\documentclass[aps,prd,twocolumn,amssymb,superscriptaddress,floatfix]{revtex4-2}
\usepackage{mathtools} 
\usepackage{graphicx}
\usepackage{epstopdf}
\usepackage{amsmath}
\usepackage{bm}
\usepackage{amssymb}
\usepackage{mathrsfs}
\usepackage{amsfonts}
\usepackage{braket}
\usepackage{setspace}
\usepackage[caption=false]{subfig}
\newcommand{\dd}[1]{\mathrm{d}#1}
\newcommand{\ii}{\mathrm{i}}
\newcommand{\E}{\mathbf E}
\newcommand{\B}{\mathbf B}
\usepackage[dvipsnames]{xcolor}

\begin{document}

\title{Searching for Ultra-Light Axions with Twisted Cavity Resonators of Anyon Rotational Symmetry with Bulk Modes of Non-Zero Helicity} 
	
\author{J. F. Bourhill}
\affiliation{Quantum Technologies and Dark Matter Research Lab, Department of Physics, University of Western Australia, Crawley, WA 6009, Australia.}
	
\author{E. C. I. Paterson}
\affiliation{Quantum Technologies and Dark Matter Research Lab, Department of Physics, University of Western Australia, Crawley, WA 6009, Australia.}
	
\author{M. Goryachev}
\affiliation{Quantum Technologies and Dark Matter Research Lab, Department of Physics, University of Western Australia, Crawley, WA 6009, Australia.}
	
\author{M. E. Tobar}
\email{michael.tobar@uwa.edu.au}
\affiliation{Quantum Technologies and Dark Matter Research Lab, Department of Physics, University of Western Australia, Crawley, WA 6009, Australia.}

\date{\today}
	
\begin{abstract}
Möbius-ring resonators stem from a well-studied geometrical structure with a one-sided topology, the Möbius strip, and have been shown to exhibit fermion rotational symmetry with respect to an untwisted ring resonator, which exhibits boson rotational symmetry \cite{Ballon2009}. Here, we present twisted hollow resonators with equilateral triangular cross-sections, leading to cavities with anyon rotational symmetry. Unlike conventional cavity resonators, these structures permit bulk resonant modes with a non-zero complex electromagnetic overlap $\mathcal C_p=\int_V\mathbf{E}_p\cdot\mathbf{B}_p^*\,d\tau$. The real and imaginary parts of this overlap couple to different observable quadratures. Although the negative-frequency component of a real axion field is immaterial for haloscopes employing a static DC magnetic field, it must be retained for a pumped resonator because the complex forms of Maxwell's equations differ for positive and negative frequencies and produce distinct upper and lower mixing branches. Retaining both branches shows that the real overlap produces amplitude-modulated sidebands, whereas the imaginary helicity of the modes studied here produces phase-modulated sidebands through the axion-photon chiral anomaly. A sensitive ultra-light dark-matter experiment may therefore be realised with an ultra-stable oscillator and a phase-sensitive readout. The sensitivity curves calculated here require the assumed noise floor to be interpreted as phase-modulation noise rather than amplitude-modulation noise, and consequently require sufficiently low residual oscillator phase noise at the axion Fourier frequency. The scheme removes the usual requirement for an external magnetic field and permits superconducting resonators with reduced surface loss and enhanced axion sensitivity.

\end{abstract}
	
\maketitle 

\section{Introduction}
	
Chirality is a fundamental property present in many physical systems, which exhibit asymmetry, ranging from particle physics \cite{Schw51,Adler69,Gooth2017vg}, topological and quantum systems \cite{Ren:2022vd,Vu:2021wl,Gooth:2019np,Pikulin16,Wang13,Wieder:2022um,Fomin:2022aa}, complex molecules, and chiroptical phenomena \cite{MacKenzie:2021vs,Tang2011,Torsi:2008wr,Hendry:2010ug,Cohen2010,Mun:2020ue}. Many of these occur as surface states, at high energy or frequency, or due to complex meta-structures \cite{Liu:2014ur,Khanikaev:2013vy,Goryachev2016} or plasmonic systems \cite{Yang16,Zhang:2020vy}, which inevitably add loss. Here, we realise a new class of electromagnetic resonator, which we call the ``anyon cavity resonator", based on twisted waveguide and ``M{\"o}bius-like" structures \cite{Ballon2009}, of twist angle $\phi$, which exhibit monochromatic bulk chiral modes at microwave frequencies with non-zero helicity through a photonic magneto-electric coupling in vacuo. The imaginary helicity of the modes studied here couples axion dark matter to the phase quadrature of a single pumped mode, without the need for a background magnetic field, with near-unity form factors in the ultra-light mass range from $10^{-22}$ to $10^{-14}$ eV \cite{ULACMB2017,Marsh17,Zhang_2018,Fedderke19}. A complementary real electric--magnetic overlap would couple to the amplitude quadrature. This not only reduces the experimental complexity but also permits low-loss superconducting resonators \cite{Creedon16,Romanenko20,Posen20,McA21} to search for ultra-light dark matter (ULDM) axions.

\begin{figure}[b!]
	\includegraphics[width=0.83\columnwidth]{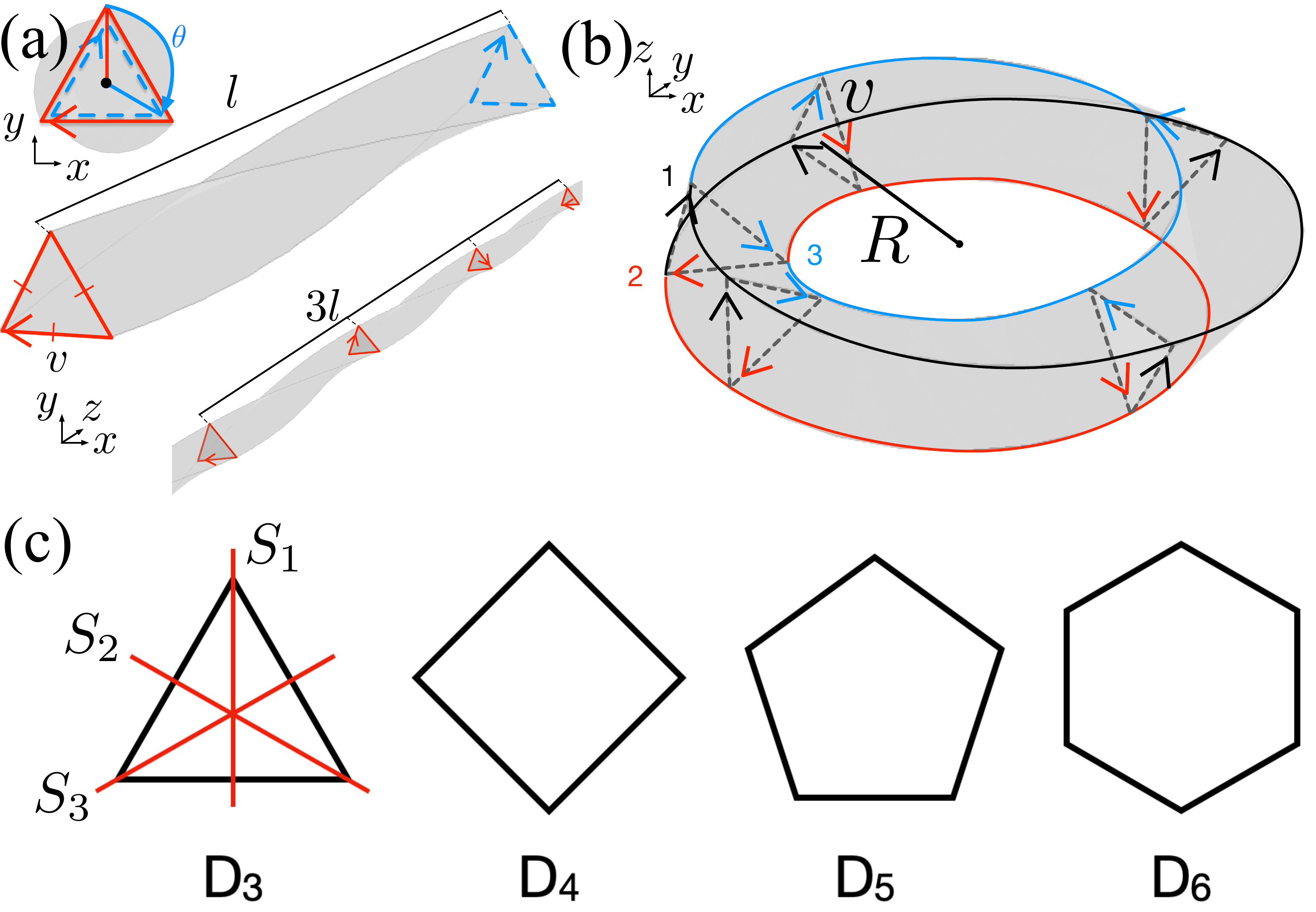}
	\caption{Geometries considered; (a) the twisted triangular waveguide resonator and (b) the triangular M\"obius-like ring resonator. (c) The dihedral group of regular convex polygons, $D_n$, which have $2n$ symmetries; $n$ rotational and $n$ reflection.}
	\label{fig:resonators}
	\end{figure}

ULDM axions have been shown to solve the standard model strong charge-parity problem \cite{PQ1977,PQ1977b,Weinberg1978,Wilczek1978},  could account for the entire dark matter of the universe \cite{Di-Luzio:2021wu,Di_Luzio_2021,Sokolov:2021uv,Visinelli19}, and are usually searched for using putative axion interactions with gluons and neutrons \cite{Wu19,Roussy21,Jiang:2021wn,Bloch22}. These experiments must be maintained for multiple years to search for ULDM, for example a particle mass of $10^{-22}$ eV corresponds to a frequency of $24.2$ nHz with a period of 1.3 years. Upconversion techniques, which utilise two real photon modes with non-zero overlap integral, $\int \mathbf{E}_n\cdot\mathbf{B}_m~d\tau$ can work in a similar way to the ULDM frequency comparison experiments if they are made near degenerate \cite{Goryachev2019,Thomson:2021wk,Lasenby2020,Lasenby2020b,berlin2020axion,Cat21,ABerlin2021,UpconvCat23}. However, feedthrough, mode coupling, non-linear effects, and injection locking are likely to limit the minimum frequency detuning between two modes, so they are not configurable for ultra-light axions. In contrast, the anyon cavity resonator has eigenmodes of non-zero helicity \cite{Cameron_2012,Cameron_2012b,Bliokh_2013,Bliokh14,Bliokh_2016,BLIOKH20151,Alpeggiani18,Wei2020}, where the sign of the helicity depends on whether the modes $\mathbf{E}$ field and $\mathbf{B}$ field have net parallel alignment ($+$ve) or net anti-parallel alignment ($-$ve). The electromagnetic helicity $\mathscr H$ can be defined as the normalised imaginary component of the field overlap integral $\mathcal C_p$:

\begin{align}
\mathcal C_p&=\int_V\mathbf E_p(\vec r)\cdot\mathbf B_p^*(\vec r)\,d\tau
=C_R+\ii C_I,
\label{eq:complexOverlap}\\
U_E&=\frac{\epsilon_0}{4}\int_V|\mathbf E_p|^2d\tau,
\qquad
U_B=\frac{1}{4\mu_0}\int_V|\mathbf B_p|^2d\tau,\nonumber\\
U_p&=U_E+U_B,
\label{eqenergies}\\
\frac{\epsilon_0c}{U_p}\mathcal C_p&=\mathscr N+\ii\mathscr H,
\label{eqhelicities}\\
\mathscr N&=\frac{\epsilon_0c}{U_p}C_R,
\qquad
\mathscr H=\frac{\epsilon_0c}{U_p}C_I.\nonumber
\end{align}
At resonance, $U_E=U_B=U_p/2$, and we can therefore say that $U_p=2\sqrt{U_B}\sqrt{U_E}$. We can then write the real and imaginary overlap terms as
\begin{align}
\mathscr N =& \frac{2C_R}{\sqrt{\int_V|\mathbf E_p|^2d\tau\int_V|\mathbf B_p|^2d\tau}},\\
\mathscr H =& \frac{2C_I}{\sqrt{\int_V|\mathbf E_p|^2d\tau\int_V|\mathbf B_p|^2d\tau}}.
\label{eq:oldformhelicity}
\end{align}
It will be shown that presence of a real overlap $\mathscr N$ term results in an interaction with an axion field producing amplitude modulation (AM) of the mode, while the imaginary overlap $\mathscr H$ is the normalised electromagnetic helicity and will result in an interaction with an axion field producing phase modulation (PM). 

Electromagnetic helicity is tied to the conservation of Zilch \cite{Lipkin:1964wb}, or photonic chirality \cite{Cohen2010}, and is usually considered when photons interact with chiral molecules or materials \cite{Cohen2010,Tang2011,Hendry:2010ug}. Analogously, the axion couples to photons through a chiral anomaly, whether in condensed matter physics \cite{Wang13,Gooth:2019np} or in quantum chromodynamics (QCD) \cite{PQ1977,Peccei2006,Weinberg1978,Wilczek1978}; introducing the axion therefore modifies the electromagnetic equations of motion by a chiral interaction.

\section{Twisted Triangular Resonators}
We investigate two twisted triangular cavity configurations as depicted in Fig.\ref{fig:resonators}: (a) a linear twisted resonator with electrically conducting boundary conditions at the ends, which allows arbitrary twist angle, $\phi$, and (b) a twisted ring resonator similar to a M\"obius strip, with only discrete allowed $\phi$ values ($\phi=\mathbb{Z}\pi/3$). 

It has been previously demonstrated that a M\"obius ring resonator with rectangular cross-section and $\pi$ twist angle exhibits fermion rotational symmetry with respect to the boson rotational symmetry of a ring resonator with no twist angle \cite{Ballon2009}. This can be understood by considering some 2D function confined to the resonator's cross-section; in the ring resonator, the function is invariant under one full cycle of the cavity circumference, whilst in the M\"obius configuration it requires two full cycles. In the triangular ring system depicted in Fig.\ref{fig:resonators}(b) we observe that such a 2D function confined to the cavity cross-section (the arrow head) is invariant after three rotations. Futhermore, considering the linear case of Fig.\ref{fig:resonators}(a), we are free to set the twist angle $\phi$ to any value, and therefore some 2D function will be invariant after $2\pi/\phi$ cycles of the cavity (3 lengths are required for invariance in the depicted case). Thus we can state that such systems exhibit anyon rotational symmetry.

\subsection{Twisted Triangular Linear Resonator}
In the linear resonator of Fig.\ref{fig:resonators}(a), by introducing the twist angle $\phi$ the resonator's mirror symmetry is broken and the resulting geometry is chiral; right handed for $\phi>0$ and left handed for $\phi<0$. This geometry dictates the boundary conditions which are used to solve Maxwell{'}s equations for the resonant electromagnetic modes within this volume.  The introduced mirror asymmetry of the boundary conditions results in a magneto-electric coupling \cite{electromagneticchirality}, which acts to couple the degenerate TE and TM modes of the untwisted cavity together into a new orthogonality basis. This can be alternatively interpreted as a dual transformation \cite{Alpeggiani18} of the no-longer orthogonal $\mathbf{E}$ and $\mathbf{B}$ fields into a new orthogonality basis $\mathbf{E^\prime}$ and $\mathbf{B^\prime}$ which are some combination of the original fields.

The result is that as a function of twist angle, the new eigenmodes of the cavity have some non-zero $\mathbf{E}.\mathbf{B}$ product and hence $\mathscr{H}_p$, and are asymmetrically detuned in opposite directions from their untwisted frequencies. The origin of this mode tuning can be thought of as a normal mode splitting between two coupled harmonic oscillators and the sign of the resulting helicity reflects the phase with which they're coupling. 
	\begin{figure}[b!]
	\includegraphics[width=0.82\columnwidth]{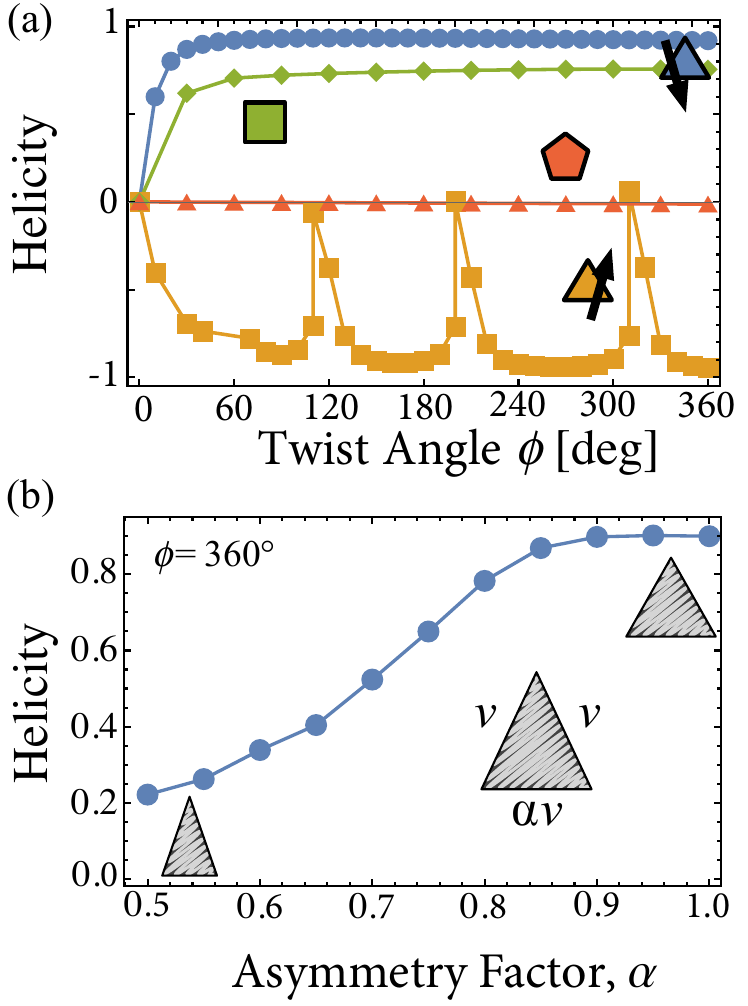}
	\caption{(a) Calculated $\mathscr{H}_p$ values as a function of $\phi$ for right and left handed modes, with equilateral triangular (blue and yellow), square (green) and pentagonal (red) cross-section. Modes tuning up in frequency have -ve $\mathscr{H}_p$ (with spurious mode interactions), while modes tuning down have -ve as indicated by the arrow.(b) Impact of varying the equilateral symmetry of teh cross-section $\alpha$ on mode helicities for $\phi= 360^\circ$.}
	\label{fig:vertices}
	\end{figure}
When considering the emergent helicity of the modes when twisted, there are two impacting factors of the cavity cross section - the polygon order and the symmetry. In regards to the former, the equilateral triangle is the lowest order polygon of the dihedral group of regular polygons (see Fig. \ref{fig:resonators}(c)) and as such, least approximates a circle. One can imagine the two resonators depicted in Fig.\ref{fig:resonators}(a) and (b) with a circular cross-section; they would not appear any different from a cylinder or torus cavity, respectively. Therefore it reasons that the greatest impact of the twisting angle $\phi$ will be generated by the lowest order regular polygon. 

This is further demonstrated in Fig.\ref{fig:vertices}(a) as the greatest magnitude helicity modes in cavities with D$_3$ to D$_5$ polygon cross-sections are shown as a function of twist angle. The two D$_3$ modes (one that tunes up and one that tunes down in frequency) approach near unity helicity, whilst higher order cross-sections result in smaller values.

The symmetry of the cross-section also plays an important role, as a perfectly symmetric cross section (i.e. equilateral trianlge, square etc.) results in TE and TM modes being tuned coincident in frequency with one another. Just like any coupled mode system,  hybridisation of the two individual systems is greatest when their frequencies are made degenerate. This is true for the helicity of the hybrid mode here, as demonstrated by Fig.\ref{fig:vertices}(b), in which one of the vertex lengths is multiplied by some factor $\alpha$, and we observe that helicity is maximised for the equilateral case. 
	\begin{figure}[b!]
	\includegraphics[width=1.0\columnwidth]{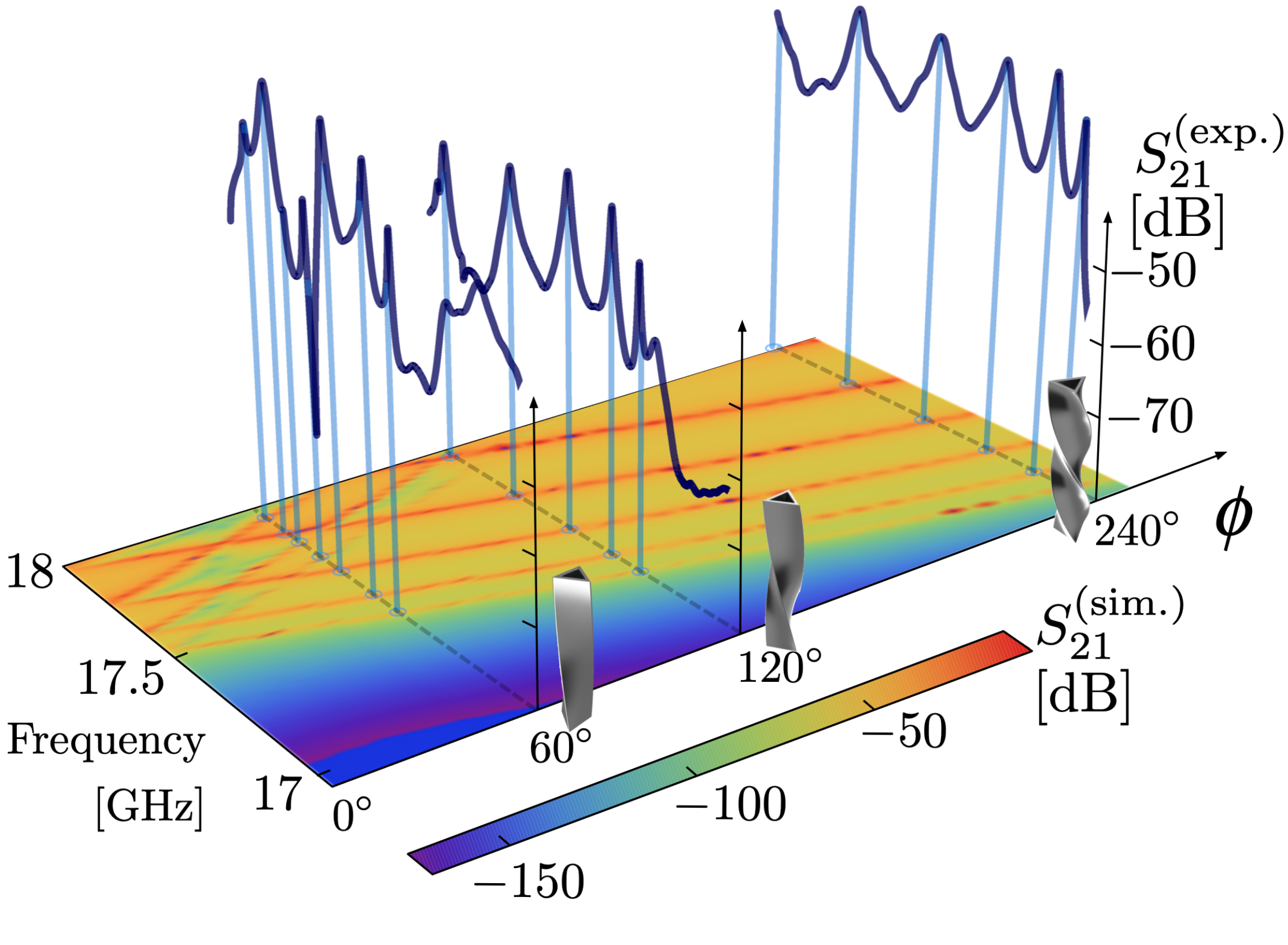}
	\caption{Experimental transmission measurements,  $S_{21}$, of the $60^\circ$, $120^\circ$ and $240^\circ$ twisted triangular waveguide cavities (dark blue) overlaying simulated frequency response, shown as the colour density plot. Light blue vertical lines indicate the corresponding locations of the resonant transmission peaks of the experimental data.}
	\label{fig:Exp}
	\end{figure}
\begin{figure}[b!]
\includegraphics[width=1.0\columnwidth]{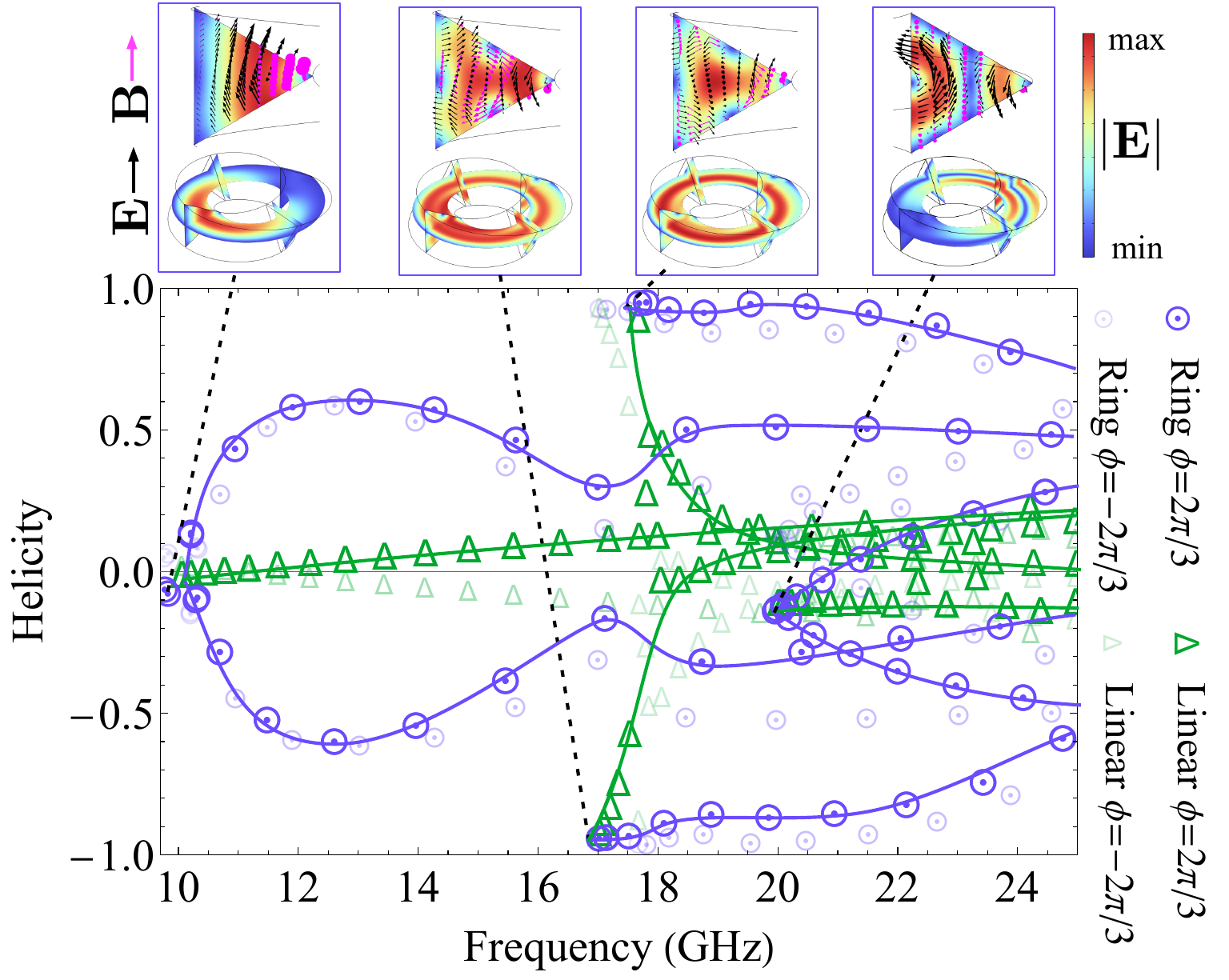}
\caption{Simulated $\mathscr{H}_p$ values for eigenmodes in an equilateral triangular cross-section ring resonator with $R=150/2\pi~\text{mm}$, $v=20$ mm and $\theta=120^\circ$ (purple circles) compared to the same dimension equivalent linear resonator (green triangles). Purple and green lines link modes of a common family.  For the first mode fo each mode family, $|\vec{E}|$ is plotted in the inset colour density plots and $\mathbf{E}$ and $\mathbf{B}$ fields as arrows. The opposite twist ($\phi<0$) data is also plotted, demonstrating a reflection of helicity values in the horizontal axis.}
\label{fig:ring}
\end{figure}
To confirm the expected frequency splitting in the linear resonator as $\phi$ is varied, we 3D printed three aluminium prototypes with right handed twist ($+ve~\phi$) using selective laser melting,  using a similar method to previously reported superconducting cylindrical cavity resonators \cite{Creedon16}. These resonators have a length of 150 mm and their cross section has a vertex length of 20 mm.  The experimentally obtained transmission spectra are measured at room temperature through co-axial probes placed at the cavity end faces and are plotted in the z-axis of Fig.\ref{fig:Exp}, which sit above the spectra obtained from finite element modelling (FEM). We note exceptional agreement between the frequency locations of the resonant modes and the predcited tuning from the simulations.  The lowest order mode with high helicity has a bandwidth of $\kappa\approx4.1$ MHz, invariant of $\phi$, consistent with a surface resistance of $R_s\sim470$ m$\Omega$ at room temperature. 

\subsection{Twisted Triangular Ring Resonator}

Next we consider the case of bending the two end faces of the resonator around and forming a ring as shown in Fig.\ref{fig:resonators}(b); removing two metallic boundaries from the resonator and therefore allowing modes with lower loss.  In addition to the resonant frequencies, the electromangetic helicity and geometry factor \cite{1393215}, $G_p=Q_{0p}R_s$, where $Q_{0p}$ is the unloaded quality factor of the modes, can be obtained from the FEM. The results are presented in Tab.\ref{tab:GF} for the greatest negative and positive helicity modes, and demonstrate a factor of 3 improvement in $Q$-factors for the ring resonator compared to the linear version. However, our attempts to 3D print a ring resonator require two seperate parts to be made, which require a knife edge for optimum electrical connectivity, which will need to be considered carefully to attain the predicted Q-factors. 

\begin{table}[htbp]
\centering
		\begin{tabular}{|c|c|c|c|}
\hline
Resonator  & $f_p$ (GHz) & $G_p$ ($\Omega$) & $\mathscr{H}_p$\\
\hline
Linear  & 17.044 & 1950 & -0.931\\
Linear & 17.688 & 1920 & 0.8796\\
\hline
Ring & 17.022 & 6200 & -0.931\\ 
Ring & 17.814 & 7290 & 0.954\\
\hline
\end{tabular}
\caption{Simulated $f_p$, $G_p$ and $\mathscr{H}_p$ values for the greatest $|\mathscr{H}_p|$ modes for the linear and ring resonators with $l=150$ mm, $\nu=20$ mm, $\alpha=1$ and $\theta=120^\circ$.}
\label{tab:GF}
\end{table}
For comparison, the TE$_{011}$ mode in a cylindrical resonator has a G-factor of the order 700 and the TM$_{010}$ mode is of order 300 \cite{Creedon16}, so $Q$-factors in the anyon cavity resonator, if limited by material surface resistance will have approximately an order of magnitude larger $Q$-factors. Recent results have shown that an $R_s$ of 0.63 n$\Omega$ at 1.3 GHz is achievable \cite{Romanenko20,Posen20}, which would result in $Q$-factors as high as $10^{13}$ if the best niobium could be implemented as a ring resonator. The 3D printed aluminium has previously achieved an $R_s$ of 0.2 m$\Omega$ at 10 GHz at mK temperatures \cite{Creedon16}, which would mean a $Q$-factor of $10^7$ is initially possible.  For axion detection, a lower frequency obtainable with increased area of cross section and volume,  and a lower $R_s$ will enhance the sensitivity.

A comparison of the helicities of the resonant modes in the linear and ring cases are presented in Fig.\ref{fig:ring}.  It can be observed that the greatest $|\mathscr{H}_p|\sim1$ modes occur at the same frequency for both resonators, which is the cutoff frequency for the TE$_{11}$ and TM$_{11}$ mode families \cite{triangle1,triangle2}.  The key difference between the two resonators is seen for this mode branch: it is only the first few modes in the linear case which have large $|\mathscr{H}_p|$, whilst there are numerous high helicity mode candidates in the ring case.  This is a result of the lack of end-face boundary conditions permitting both $\sin$ and $\cos$ azimuthal field dependences in the ring case,  one of which is forbidden in the linear case. In the latter, as the number of axial nodes increases, the field overlap between the $\sin$ (TE) and $\cos$ (TM) modes is largest for the first few modes and then drops, whilst in the ring, the field overlap integral can be maintained as both $\sin$ and $\cos$ solutions are permitted for both original polarisations.  In both forms of the resonator, the lower frequency modes with cutoff frequency $\sim10$ GHz have no TM counterpart and therefore do not achieve unity $|\mathscr{H}_p|$.

\section{Axion-modified electrodynamics}

The axion--photon interaction follows from the chiral anomaly and is described by \cite{Sikivie2021}
\begin{equation}
\mathcal{L}_{a\gamma\gamma}
=\frac{g_{a\gamma\gamma}}{4}aF_{\mu\nu}\widetilde F^{\mu\nu}
=g_{a\gamma\gamma}a\,\mathbf E\cdot\mathbf B.
\label{AxInt}
\end{equation}
Writing the dimensionless axion angle as $\Theta(t)=g_{a\gamma\gamma}a(t)$, the modified Maxwell equations are \cite{Sikivie:1984vr,Wilczek:1987aa}
\begin{equation}
\begin{aligned}
&\nabla\cdot\left(\epsilon_0\mathbf E-\epsilon_0c\Theta\mathbf B\right)=\rho_e,\\
&\nabla\times\left(\frac{\mathbf B}{\mu_0}+\frac{\Theta\mathbf E}{\mu_0c}\right)
-\partial_t\left(\epsilon_0\mathbf E-\epsilon_0c\Theta\mathbf B\right)=\mathbf J_e,\\
&\nabla\cdot\mathbf B=0,\qquad
\nabla\times\mathbf E+\partial_t\mathbf B=0.
\end{aligned}
\label{EDEoM}
\end{equation}
Equation~\eqref{AxInt} is therefore represented in the field equations by the terms proportional to $\Theta$ in Eq.~\eqref{EDEoM}.
For nonrelativistic cold dark matter the spatial gradient of the local axion field may be neglected.  With the phasor convention $e^{-\ii\omega t}$, the real field must retain both conjugate Fourier components,
\begin{equation}
\Theta(t)=\frac12\left(\widetilde\theta e^{-\ii\omega_at}
+\widetilde\theta^*e^{+\ii\omega_at}\right).
\label{eq:theta}
\end{equation}
For a coherent tone, Eq.~\eqref{eq:theta} becomes $\Theta(t)=\theta_0\cos(\omega_at+\phi_a)$, where $\omega_a=m_ac^2/\hbar$ and $\theta_0=g_{a\gamma\gamma}a_0\ll1$.

\begin{figure}[t!]
\includegraphics[width=0.8\columnwidth]{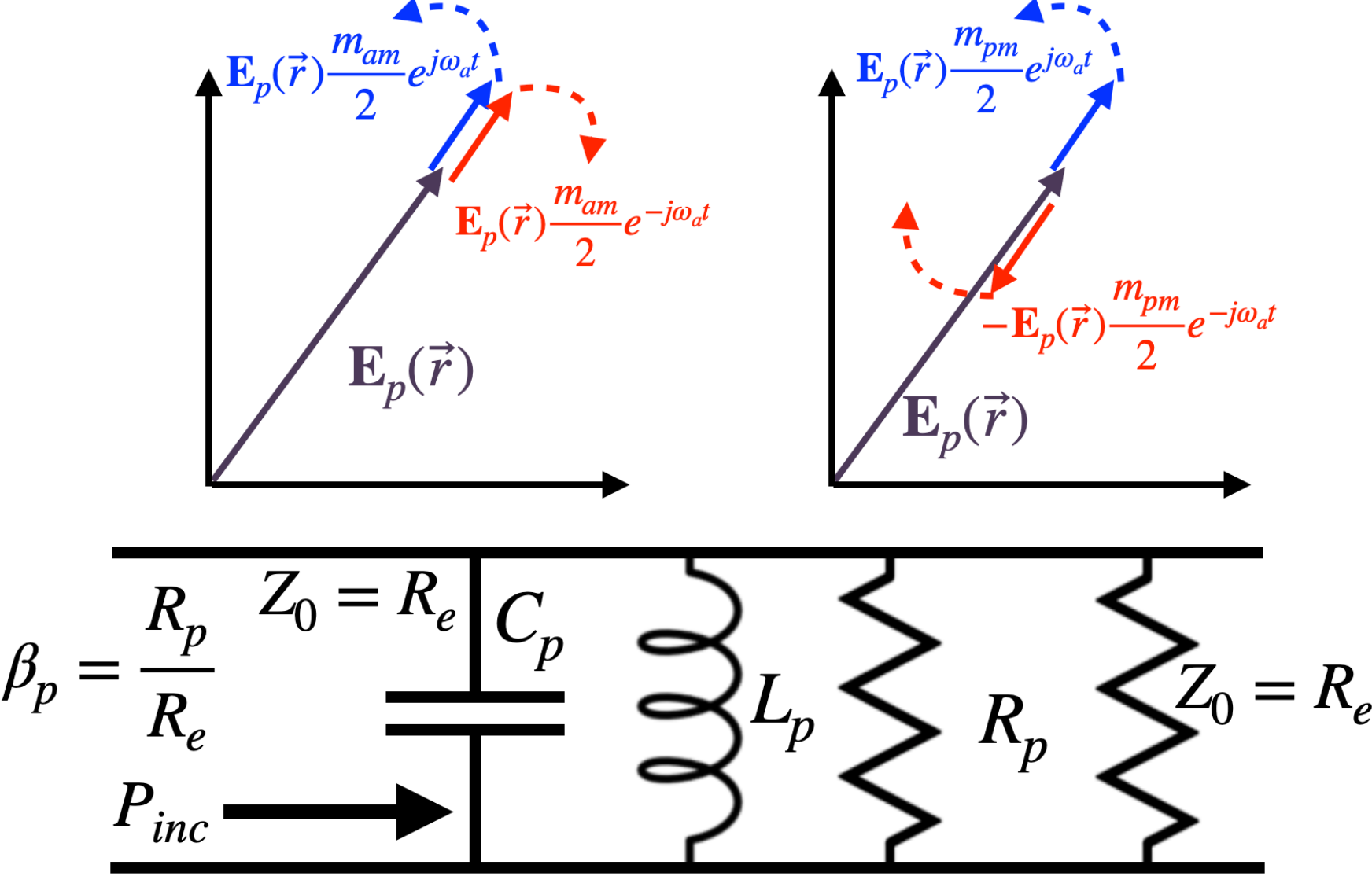}
\caption{Phasor diagram of the time-dependent vector phasor $\widetilde{\mathbf E}(\vec r,t)$ in the frame rotating at $\omega_p$. Top left: amplitude modulation. Top right: phase modulation. The single-tone modulation indices satisfy $m_{\rm AM},m_{\rm PM}\ll1$. The lower sideband is shown in red and the upper sideband in blue. Bottom: equivalent parallel LCR model of a resonant mode with coupling $\beta_p$, impedance matched when $\beta_p=1$.}
\label{phasor}
\end{figure}
The relative sideband phases associated with AM and PM are illustrated in
Fig.~\ref{phasor}.

\subsection{Positive- and negative-frequency mixing branches}

To first order in the real axion field of Eq.~\eqref{eq:theta}, the effective
current is $\mathbf J_a=-\epsilon_0c\dot\Theta\mathbf B$.  Since this source is
already first order in $\Theta$, the magnetic field in the product is the
unperturbed real carrier field
\begin{align*}
 \mathbf B^{(0)}(\vec r,t)
 &=\frac12\left[\mathbf B_p(\vec r)e^{-\ii\omega_pt}
 +\mathbf B_p^*(\vec r)e^{+\ii\omega_pt}\right],\\
 \dot\Theta(t)
 &=\frac12\left[-\ii\omega_a\widetilde\theta e^{-\ii\omega_at}
 +\ii\omega_a\widetilde\theta^*e^{+\ii\omega_at}\right].
\end{align*}
Replacing $\mathbf B^{(0)}$ by an axion-induced sideband field in this product
would generate only terms of order $\Theta^2$.  Multiplying the two real fields
above gives four Fourier components.  The two positive-frequency components
near the carrier, together with their negative-frequency conjugates, are
\begin{equation}
\begin{aligned}
 \mathbf J_a(\vec r,t)
 ={}&\frac{\ii\epsilon_0c\omega_a}{4}
 \widetilde\theta\mathbf B_p e^{-\ii\omega_+t}\\
 &-\frac{\ii\epsilon_0c\omega_a}{4}
 \widetilde\theta^*\mathbf B_p e^{-\ii\omega_-t}
 +\mathrm{c.c.}\\
 ={}&\frac12\left[
 \mathbf J_{a,+}e^{-\ii\omega_+t}
 +\mathbf J_{a,-}e^{-\ii\omega_-t}
 +\mathrm{c.c.}\right],
\end{aligned}
\label{eq:currents}
\end{equation}
where $\omega_+=\omega_p+\omega_a$ and
$\omega_-=\omega_p-\omega_a$.  Equation~\eqref{eq:currents} therefore
incorporates both sidebands explicitly. 
Substituting this axion-induced source current into the modified Amp\`ere equation,
\begin{equation*}
 \frac{1}{\mu_0}\nabla\times\mathbf B
 =\mathbf J_e+\epsilon_0\partial_t\mathbf E
 +\mathbf J_a,
\end{equation*}
the carrier and sideband equations must be evaluated at three different frequencies:
\begin{subequations}
\begin{align}
\frac{1}{\mu_0}\nabla\times\mathbf{B}_p &= \mathbf{J}_{e,p} - \ii\omega_p\epsilon_0\mathbf{E}_p, 
 \label{eq:AmperePump}\\
 \frac{1}{\mu_0}\nabla\times\mathbf{B}_+ &= \mathbf{J}_{e,+} - \ii\omega_+\epsilon_0\mathbf{E}_+ + \frac{\ii\epsilon_0 c\,\omega_a}{2}\,\widetilde{\theta}\,\mathbf{B}_p,
 \label{eq:sidebandAmperePlus}\\
 \frac{1}{\mu_0}\nabla\times\mathbf{B}_- &= \mathbf{J}_{e,-} - \ii\omega_-\epsilon_0\mathbf{E}_- - \frac{\ii\epsilon_0 c\,\omega_a}{2}\,\widetilde{\theta}^{*}\,\mathbf{B}_p.
 \label{eq:sidebandAmpereMinus}
\end{align}
\end{subequations}
Equations~\eqref{eq:sidebandAmperePlus}
and \eqref{eq:sidebandAmpereMinus} thus describe two different
positive-frequency fields; the lower-sideband equation is not the complex
conjugate of the upper-sideband equation, and each must be carried through the
energy balance separately.  Faraday's law in the same convention,
$\nabla\times\E_k=\ii\omega_k\B_k$ with $k\in\{p,+,-\}$, together with
Eqs.~\eqref{eq:AmperePump}--\eqref{eq:sidebandAmpereMinus}, gives a separate
local complex Poynting theorem at each frequency~\cite{tobar2021abraham}.  With
$\mathbf S_k=\tfrac12\E_k\times\B_k^*/\mu_0$,
\begin{subequations}
\begin{align}
\nabla\cdot\mathbf S_p={}&\frac{1}{2\mu_0}\left(\B_p^*\cdot(\nabla\times\E_p)-\E_p\cdot(\nabla\times\B_p^*)\right)\notag\\
={}&-\frac12\E_p\cdot\mathbf J_{e,p}^*
+\frac{\ii\omega_p}{2}\left(\frac{|\B_p|^2}{\mu_0}-\epsilon_0|\E_p|^2\right),
\label{eq:localPoyntingPump}\\
\nabla\cdot\mathbf S_+={}&-\frac12\E_+\cdot\mathbf J_{e,+}^*
+\frac{\ii\epsilon_0c\,\omega_a}{4}\,\widetilde\theta^*\,\E_+\cdot\B_p^*\notag\\
&+\frac{\ii\omega_+}{2}\left(\frac{|\B_+|^2}{\mu_0}-\epsilon_0|\E_+|^2\right),
\label{eq:localPoyntingPlus}\\
\nabla\cdot\mathbf S_-={}&-\frac12\E_-\cdot\mathbf J_{e,-}^*
-\frac{\ii\epsilon_0c\,\omega_a}{4}\,\widetilde\theta\,\E_-\cdot\B_p^*\notag\\
&+\frac{\ii\omega_-}{2}\left(\frac{|\B_-|^2}{\mu_0}-\epsilon_0|\E_-|^2\right).
\label{eq:localPoyntingMinus}
\end{align}
\end{subequations}
The axion terms in Eqs.~\eqref{eq:localPoyntingPlus} and
\eqref{eq:localPoyntingMinus} are $-\tfrac12\E_\pm\cdot\mathbf J_{a,\pm}^*$
with the current phasors of Eq.~\eqref{eq:currents}.  
Averaging over a time $T\gg1/\omega_a$ removes all products of fields at different frequencies, so
the time-averaged power balance is $\sum_k\operatorname{Re}\nabla\cdot\mathbf S_k$
and the axion enters only through the two sideband balances.

\subsection{Sideband complex-power balance and cavity envelope}

Integrating Eqs.~\eqref{eq:localPoyntingPlus} and
\eqref{eq:localPoyntingMinus} over the cavity volume gives
\begin{equation}
\begin{aligned}
 \mathcal P_{S,\pm}\equiv{}&\oint_S\mathbf S_\pm\cdot\hat{\mathbf n}\,\dd S\\
 ={}&-\frac12\int_V\E_\pm\cdot\mathbf J_{e,\pm}^*\,\dd V
 +2\ii\omega_\pm\left(U_{B,\pm}-U_{E,\pm}\right)\\
 &+P_{a,\pm},
\end{aligned}
\label{eq:sidebandPoyntingIntegral}
\end{equation}
where $U_{E,\pm}$ and $U_{B,\pm}$ are defined as in Eq.~\eqref{eqenergies}
with $\E_p,\B_p\rightarrow\E_\pm,\B_\pm$, and the axion source terms are
\begin{equation}
\begin{aligned}
 P_{a,+}&=\frac{\ii\epsilon_0c\,\omega_a}{4}\,\widetilde\theta^*
 \int_V\E_+\cdot\B_p^*\,\dd V,\\
 P_{a,-}&=-\frac{\ii\epsilon_0c\,\omega_a}{4}\,\widetilde\theta
 \int_V\E_-\cdot\B_p^*\,\dd V.
\end{aligned}
\label{eq:sidebandSources}
\end{equation}
To evaluate Eqs.~\eqref{eq:sidebandPoyntingIntegral} and
\eqref{eq:sidebandSources}, we use the fact that both sidebands are excited
in the pump mode itself.  For $\omega_a\ll\omega_p$, and with $\omega_a$ and cavity bandwidth,
$\kappa_p$ small compared with the detuning of any neighbouring mode that
overlaps the axion source current, the sideband fields share the pump-mode
profile and differ from it only by complex amplitudes $u_\pm$:
\begin{equation}
 \E_\pm=u_\pm\E_p,\qquad \B_\pm=u_\pm\B_p .
 \label{eq:sidebandAnsatz}
\end{equation}
Here $u_\pm$ is the sideband field amplitude normalized to the pump, with
$u_\pm=0$ in the absence of the axion.  Substituting
Eq.~\eqref{eq:sidebandAnsatz} into Eq.~\eqref{eq:sidebandPoyntingIntegral}
simplifies each of its terms.  First, the axion sources become
$P_{a,+}=\tfrac{\ii}{4}\epsilon_0c\,\omega_a\widetilde\theta^*u_+\mathcal C_p$
and
$P_{a,-}=-\tfrac{\ii}{4}\epsilon_0c\,\omega_a\widetilde\theta\,u_-\mathcal C_p$,
so both are set by the complex overlap of Eq.~\eqref{eq:complexOverlap}.
Second, the sideband energies become $U_\pm=|u_\pm|^2U_p$.  Third, we group
the surface-flux and dissipative terms into a single loss term,
$L_\pm\equiv\oint_S\mathbf S_\pm\cdot\hat{\mathbf n}\,\dd S
+\tfrac12\int_V\E_\pm\cdot\mathbf J_{e,\pm}^*\,\dd V$,
which is the power the sideband loses to the port and to internal
dissipation.  Its real part is $\kappa_pU_\pm$, with $\kappa_p$ the loaded
energy decay rate.  Its imaginary part vanishes for ohmic loss
($\mathbf J_{e,\pm}=\sigma\E_\pm$ with real $\sigma$).
We therefore set
$L_\pm=\kappa_pU_\pm$.  Finally, the reactive energy term reflects the fact that the sidebands are
driven off resonance.  A mode driven at its resonance stores equal electric
and magnetic energy, whereas a mode driven at detuning $\Delta$ acquires an
imbalance which, for a current-driven resonator, follows from Foster's theorem
in susceptance form~\cite{Foster1924} as $U_B-U_E=-(\Delta/\omega)U$ to first
order in $\Delta/\omega$.  With the pump on resonance, the sidebands are
detuned by $\Delta=\pm\omega_a$, giving
$2\omega_\pm(U_{B,\pm}-U_{E,\pm})=\mp2\omega_aU_\pm$.

With these substitutions, Eq.~\eqref{eq:sidebandPoyntingIntegral} becomes
\begin{equation}
 U_\pm\left(\kappa_p\pm2\ii\omega_a\right)=P_{a,\pm}.
 \label{eq:sidebandBalance}
\end{equation}
Dividing Eq.~\eqref{eq:sidebandBalance} by $2U_pu_\pm$, taking the complex conjugate, and using
Eq.~\eqref{eqhelicities} gives the sideband amplitudes
\begin{equation}
\begin{aligned}
 u_+&=-\frac{\ii\omega_a}{8}\,
 \frac{(\mathscr N-\ii\mathscr H)\,\widetilde\theta}{\kappa_p/2-\ii\omega_a},\\
 u_-&=+\frac{\ii\omega_a}{8}\,
 \frac{(\mathscr N-\ii\mathscr H)\,\widetilde\theta^*}{\kappa_p/2+\ii\omega_a}.
\end{aligned}
\label{eq:sidebandAmplitudes}
\end{equation}

Neither amplitude alone separates $\mathscr N$ from $\mathscr H$; the
separation appears in the total pump-mode field.  Because the sidebands share
the pump-mode profile, the total electric field can be written as
\begin{equation}
\begin{aligned}
 \E(\vec r,t)&=\frac12\left[\left(1+u(t)\right)\E_p(\vec r)e^{-\ii\omega_pt}
 +\mathrm{c.c.}\right],\\
 u(t)&=u_+e^{-\ii\omega_at}+u_-e^{+\ii\omega_at},
\end{aligned}
\label{eq:envelope}
\end{equation}
where $u(t)$ is the complex envelope of the pump mode.  Substituting
Eq.~\eqref{eq:sidebandAmplitudes} into $u$, one can verify directly that it
satisfies
\begin{equation}
 \dot u+\frac{\kappa_p}{2}u=\frac{\mathscr N-\ii\mathscr H}{4}\,\dot\Theta(t),
 \label{eq:envelopeEquation}
\end{equation}
which is the standard envelope equation for a driven resonant mode, with
field decay rate $\kappa_p/2$ and an axion drive proportional to
$\dot\Theta$.  The drive is a fixed complex coefficient times the real
function $\dot\Theta$ only because both sidebands are retained with opposite
source signs.  Since $\dot\Theta$ is real, the real and imaginary parts of
Eq.~\eqref{eq:envelopeEquation} decouple:
\begin{subequations}
\begin{align}
 \frac{\dd}{\dd t}\operatorname{Re}u+\frac{\kappa_p}{2}\operatorname{Re}u
 &=\frac{\mathscr N}{4}\dot\Theta(t),
 \label{eq:AMamplitudeEquation}\\
 \frac{\dd}{\dd t}\operatorname{Im}u+\frac{\kappa_p}{2}\operatorname{Im}u
 &=-\frac{\mathscr H}{4}\dot\Theta(t).
 \label{eq:PMamplitudeEquation}
\end{align}
\end{subequations}
To first order in $u$, $1+u=(1+\operatorname{Re}u)\,e^{\ii\operatorname{Im}u}$,
so Eq.~\eqref{eq:envelope} becomes
\begin{equation}
 \E(\vec r,t)=\left(1+\operatorname{Re}u\right)
 \operatorname{Re}\!\left[\E_p(\vec r)\,
 e^{-\ii(\omega_pt-\operatorname{Im}u)}\right].
 \label{eq:polarEnvelope}
\end{equation}
Equation~\eqref{eq:polarEnvelope} shows that $\operatorname{Re}u=\delta A/A_0$ is the fractional amplitude change and,
writing the carrier phase as $\omega_pt+\phi$, $\delta\phi=-\operatorname{Im}u$
is the phase change.  Thus $\mathscr N$ produces AM and $\mathscr H$ produces PM.

The Fourier component of Eq.~\eqref{eq:AMamplitudeEquation} at $\omega_a$ is
\begin{equation}
 \operatorname{Re}u(\omega_a)=
 \frac{-\ii\omega_a\mathscr N/4}
 {\kappa_p/2-\ii\omega_a}\,\Theta(\omega_a),
 \label{eq:AMtransfer}
\end{equation}
so Eq.~\eqref{eq:AMtransfer} gives, for $\Theta(t)=\theta_0\cos(\omega_at+\phi_a)$, the rms fractional AM
\begin{equation}
  |m_{\mathrm{AM,rms}}|=
 \frac{\theta_0|\mathscr N|\omega_a}
 {4\sqrt2\sqrt{(\kappa_p/2)^2+\omega_a^2}}.
 \label{eq:AMrmsDirect}
\end{equation}

Equation~\eqref{eq:PMamplitudeEquation} is the equation obeyed by
a fixed-frequency pump when the resonance shifts to $\omega_p+\delta\omega_p$,
which adds $-\ii\delta\omega_p(1+u)\approx-\ii\delta\omega_p$ to the right-hand
side of Eq.~\eqref{eq:envelopeEquation}.  The imaginary overlap is therefore
equivalent to an axion-induced frequency shift
\begin{equation}
\delta\omega_p(t)=\frac{\mathscr H}{4}\dot\Theta(t),
 \label{eq:frequencyShift}
\end{equation}
with peak deviation, for $\Theta(t)=\theta_0\cos(\omega_at+\phi_a)$,
\begin{equation}
 \delta\omega_{p,\mathrm{pk}}
 =\frac{|\mathscr H|\theta_0\omega_a}{4}.
 \label{eq:peakFrequencyShift}
\end{equation}
The sign depends on the phasor and axion-interaction conventions, but the
magnitude and reactive character do not, and Eq.~\eqref{eq:frequencyShift}
contains no intrinsic $Q_p$ enhancement.  In the experiment proposed here,
the fixed-frequency pump uses the cavity as a phase
discriminator, and the Fourier component of
Eq.~\eqref{eq:PMamplitudeEquation} at $\omega_a$ gives
\begin{equation}
 \delta\phi(\omega_a)=
 \frac{2Q_p/\omega_p}{1-2\ii Q_p\omega_a/\omega_p}
 \delta\omega_p(\omega_a).
 \label{eq:phaseTransfer}
\end{equation}
Thus the intrinsic frequency shift in Eq.~\eqref{eq:frequencyShift} is
converted into a measured PM signal by the cavity susceptibility.

\subsection{Modulation indices, spectra, and signal-to-noise ratios}

Define the cavity response denominator, $D_a=1+4Q_p^2\left(\frac{\omega_a}{\omega_p}\right)^2$, which is used in the original paper to take into account the finite cavity resonator bandwidth via the substitution $Q_p\rightarrow Q_p/\sqrt{D_a}$.  If $\theta_0$ is the peak axion angle, the AM result obtained from Eq.~\eqref{eq:AMamplitudeEquation} and the PM result obtained from Eqs.~\eqref{eq:peakFrequencyShift} and \eqref{eq:phaseTransfer} give the corrected rms modulation indices
\begin{align}
 |m_{\mathrm{AM,rms}}|&=
 \frac{\theta_0|\mathscr{N}|}{2\sqrt2} \frac{Q_p\omega_a/\omega_p}{\sqrt{D_a}},
 \label{eq:mam}\\
 |m_{\mathrm{PM,rms}}|&=
 \frac{\theta_0|\mathscr{H}|}{2\sqrt2} \frac{Q_p\omega_a/\omega_p}{\sqrt{D_a}}.
 \label{eq:mpm}
\end{align}
Equation~\eqref{eq:mam} is exactly Eq.~\eqref{eq:AMrmsDirect} rewritten using
$\kappa_p=\omega_p/Q_p$ and
$\sqrt{(\kappa_p/2)^2+\omega_a^2}=(\omega_p/2Q_p)\sqrt{D_a}$.
Thus the AM follows directly from the real part of the envelope equation, Eq.~\eqref{eq:envelopeEquation}, whereas Eq.~\eqref{eq:mpm} contains the phase-discriminator response $Q_p\omega_a/(\omega_p\sqrt{D_a})$ to the intrinsic frequency shift in Eq.~\eqref{eq:frequencyShift}.

With coupling coefficient $\beta_p$ and incident carrier power
$P_{\rm inc}$, the total powers in the two modulation sidebands are
\begin{subequations}
\begin{align}
 P_{\mathrm{AM}}&=
 Q_p^2\frac{4\beta_p^2}{(1+\beta_p)^2}
 \frac{(\omega_a/\omega_p)^2}{D_a}
 \frac{\theta_0^2\mathscr{N}^2}{8}P_{\rm inc},
 \label{eq:powerAM}\\
 P_{\mathrm{PM}}&=
 Q_p^2\frac{4\beta_p^2}{(1+\beta_p)^2}
 \frac{(\omega_a/\omega_p)^2}{D_a}
 \frac{\theta_0^2\mathscr{H}^2}{8}P_{\rm inc}.
 \label{eq:powerPM}
\end{align}
\end{subequations}
Equations~\eqref{eq:powerAM} and \eqref{eq:powerPM} are the total powers in the two modulation sidebands.  To state the spectral form without dimensional ambiguity, define the axion field $a(t)$ by
$\Theta(t)=g_{a\gamma\gamma}a(t)$ and define the one-sided spectra by
$S_\Theta(f)=g_{a\gamma\gamma}^2S_a(f)$.  Thus $S_\Theta$, the fractional-amplitude-noise spectrum $S_{am}$, and the phase-noise spectrum $S_{pm}$ have SI spectral units $\mathrm{Hz}^{-1}$.  The observable spectra separate into
\begin{subequations}
\begin{align}
 S_{A,\mathrm{AM}}(f)&=g_{a\gamma\gamma}^2
 \frac{\beta_p^2}{2(1+\beta_p)^2}
 \frac{Q_p^2}{D_a}\left(\frac{f_a}{f_p}\right)^2
 \mathscr{N}^2 S_a(f),
 \label{eq:spectrumAM}\\
 S_{A,\mathrm{PM}}(f)&=g_{a\gamma\gamma}^2
 \frac{\beta_p^2}{2(1+\beta_p)^2}
 \frac{Q_p^2}{D_a}\left(\frac{f_a}{f_p}\right)^2
 \mathscr{H}^2 S_a(f).
 \label{eq:spectrumPM}
\end{align}
\end{subequations}

Equations~\eqref{eq:spectrumAM} and \eqref{eq:spectrumPM} must be compared with the noise spectrum of the quadrature being measured.  For an integration time longer than the axion coherence time, the two signal-to-noise ratios are
\begin{align}
 \mathrm{SNR}_{\mathrm{AM}}&=
 \frac{g_{a\gamma\gamma}\beta_p|\mathscr{N}|}{\sqrt2(1+\beta_p)}
 \frac{Q_p}{\sqrt{D_a}}
 \left(\frac{10^6t}{\omega_a}\right)^{1/4}
 \frac{\sqrt{\rho_ac^3}}{\omega_p\sqrt{S_{am}(f_a)}},
 \label{eq:snram}\\
 \mathrm{SNR}_{\mathrm{PM}}&=
 \frac{g_{a\gamma\gamma}\beta_p|\mathscr{H}|}{\sqrt2(1+\beta_p)}
 \frac{Q_p}{\sqrt{D_a}}
 \left(\frac{10^6t}{\omega_a}\right)^{1/4}
 \frac{\sqrt{\rho_ac^3}}{\omega_p\sqrt{S_{pm}(f_a)}}.
 \label{eq:snrpm}
\end{align}
Equations~\eqref{eq:snram} and \eqref{eq:snrpm} use the one-sided fractional amplitude-noise and phase-noise power spectral densities $S_{am}$ and $S_{pm}$. 

\subsection{Sensitivity to Axions in Quantum Electromagnetodynamics}

Recently it was shown that there is a further modification of axion electrodynamics through the existence of putative magnetically charged matter \cite{SokolovMonopole22,sokolov2023generic}.  Assuming $\nabla a=0$, the modified Amp\`ere and Faraday laws may be written in phasor-amplitude form \cite{TobarQEMD22,tobar2023searching}.
The positive- and negative-frequency components of the real axion field must
again be retained separately.  To first order in $a$, their products with the
positive-frequency pump generate the upper and lower sidebands, respectively.
The corresponding equations at $\omega_+=\omega_p+\omega_a$ and
$\omega_-=\omega_p-\omega_a$ are
\begin{subequations}
\label{Farmon}
\begin{align}
\frac{1}{\mu_0}\nabla\times\mathbf B_+
={}&\mathbf J_{e,+}-\ii\omega_+\epsilon_0\mathbf E_+\notag\\
&+\frac{\ii\omega_a\epsilon_0}{2}\widetilde a
 \left(cg_{a\gamma\gamma}\mathbf B_p-g_{aEM}\mathbf E_p\right),
\label{eq:QEMDAmperePlus}\\
\frac{1}{\mu_0}\nabla\times\mathbf B_-
={}&\mathbf J_{e,-}-\ii\omega_-\epsilon_0\mathbf E_-\notag\\
&-\frac{\ii\omega_a\epsilon_0}{2}\widetilde a^*
 \left(cg_{a\gamma\gamma}\mathbf B_p-g_{aEM}\mathbf E_p\right),
\label{eq:QEMDAmpereMinus}\\
\nabla\times\mathbf E_+
={}&\ii\omega_+\mathbf B_+\notag\\
&+\frac{\ii\omega_a}{2}\widetilde a
 \left(\frac{g_{aMM}}{c}\mathbf E_p-g_{aEM}\mathbf B_p\right),
\label{eq:QEMDFaradayPlus}\\
\nabla\times\mathbf E_-
={}&\ii\omega_-\mathbf B_-\notag\\
&-\frac{\ii\omega_a}{2}\widetilde a^*
 \left(\frac{g_{aMM}}{c}\mathbf E_p-g_{aEM}\mathbf B_p\right).
\label{eq:QEMDFaradayMinus}
\end{align}
\end{subequations}
Here $g_{aEM}$ and $g_{aMM}$ are additional axion--photon couplings if
high-energy magnetic charge exists.  The factors of $1/2$ follow from the
positive- and negative-frequency decomposition of the real axion field, and
the opposite source signs follow from
$\partial_t(\widetilde a e^{-\ii\omega_at})=-\ii\omega_a\widetilde a
e^{-\ii\omega_at}$ and
$\partial_t(\widetilde a^*e^{+\ii\omega_at})=+\ii\omega_a\widetilde a^*
e^{+\ii\omega_at}$.  Thus the lower-sideband equations are not obtained by
simply reusing the upper-sideband phasor equations.  Equations~\eqref{Farmon}
preserve the AM/PM quadrature separation above, with the relevant axion angle
replaced by $\theta_0=(g_{a\gamma\gamma}+g_{aMM})a_0$ for this interaction.
Thus the limits presented here may be interpreted as limits on
$g_{a\gamma\gamma}+g_{aMM}$, allowing a single-mode search for ultra-light
axions in the mass range $10^{-22}$ to $10^{-14}$ eV without dual-mode
excitation or a large-volume magnet.

\section{Possible Experimental Configurations and Sensitivity}

\begin{figure}[b!]
\includegraphics[width=0.65\columnwidth]{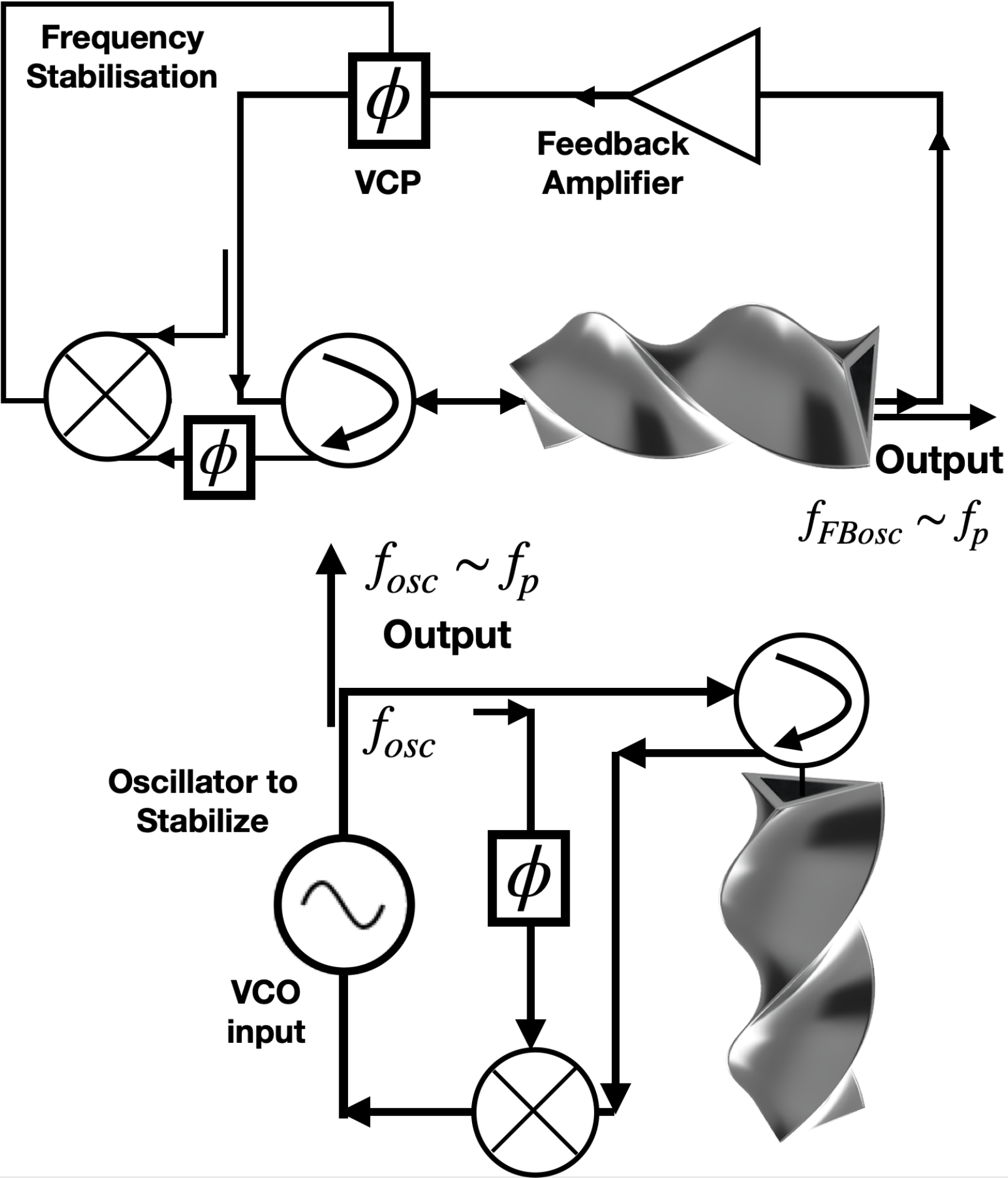}
\caption{Top, Schematic of a frequency stabilized loop oscillator configuration using a phase detector (mixer) to frequency lock the feed back oscillation, $f_{FBosc}$ to the anyon cavity resonant frequency, $f_p$, via a voltage controlled phase shifter (VCP). Bottom, schematic of an external voltage controlled oscillator (VCO) frequency locked to the anyon cavity resonator, using a phase detector (mixer) , so $f_{osc}\sim f_p$. In both cased the resonator and phase detector combine to make a frequency discriminator, which sets the oscillator frequency lock equal to the frequency of the anyon cavity resonator.}
\label{oscillator}
\end{figure}
The previously studied dual-mode upconversion experiment excites a pump and a readout mode in one cavity \cite{UpconvCat23}. Here a single mode acts as its own background field and is therefore sensitive to ultra-light axions in the upconversion limit. The appropriate implementation is a frequency-stabilized oscillator locked tightly to the axion-sensitive cavity, as shown in Fig.~\ref{oscillator}. Interferometric \cite{Ivanov2009aa,Ivanov2006,Ivanov2009,Ivanov1998} and Pound-modulated systems \cite{locke} provide phase-sensitive readout and can reduce residual oscillator phase noise. This is essential for the modes calculated in this work because $\mathscr H_p=\mathscr H$ and the axion signal is PM. Close-to-carrier phase noise is affected by the Leeson response and can be substantially larger than the oscillator AM noise. Consequently the relevant experimental noise is the measured $S_{pm}(f)$ at the axion offset, including the oscillator, locking electronics, phase detector, and readout chain. Amplitude noise remains relevant only for a mode with non-zero $\mathscr N$ or through technical AM--PM conversion.

\begin{figure}[t!]
\includegraphics[width=1.0\columnwidth]{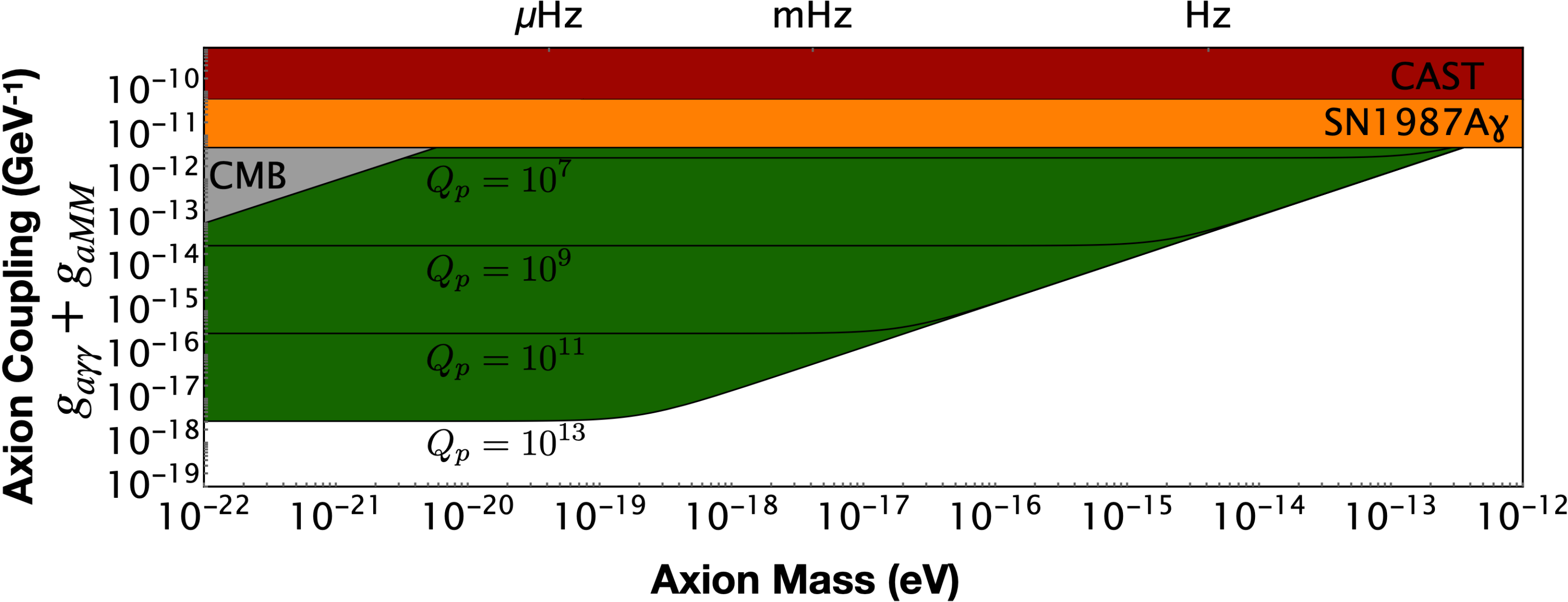}
\caption{Predicted exclusion regions of axion-photon coupling strength $g_{a\gamma\gamma}+g_{aMM}$ for a low-mass axion search with a 1.3 GHz anyon cavity and several $Q_p$ values (green), together with astrophysical limits. The curves are valid when the assumed $-160$ dBc/Hz oscillator-noise floor is interpreted as residual PM noise $S_{pm}(f_a)$. Current 3D-printed aluminium technology can reach $Q_p=10^7$ \cite{Creedon16}; the upper range represents prospective superconducting technology \cite{Romanenko20,Posen20,McA21}.}
\label{fig:exclusion}
\end{figure}

The sensitivity estimates are shown in Fig.~\ref{fig:exclusion}. They are obtained by solving the PM expression, Eq.~\eqref{eq:snrpm}, for $g_{a\gamma\gamma}$ with $\mathrm{SNR}=1$, critical coupling $\beta_p=1$, $|\mathscr H|=1$, $\omega_p/(2\pi)=1$ GHz, and $\rho_a=8\times10^{-22}$ kg/m$^3$ ($0.45$ GeV/cm$^3$). The curves retain the numerical values originally calculated if the assumed $-160$ dBc/Hz floor is interpreted as $S_{pm}(f_a)$ rather than $S_{am}(f_a)$. This reinterpretation does not guarantee that the benchmark is experimentally attainable: it must be demonstrated as residual phase noise over the relevant offset-frequency range. Sensitivities for several loaded quality factors $Q_p$ are shown. With the quantum-electromagnetodynamic modification \cite{SokolovMonopole22,sokolov2023generic}, the same curves constrain $g_{a\gamma\gamma}+g_{aMM}$ \cite{TobarQEMD22,tobar2023searching}.

The calculation does not include technical systematics such as temperature, pressure, vibration, magnetic sensitivity, loop-gain fluctuations, mixer flicker, or AM--PM conversion. These effects must be characterized in the phase quadrature. Related oscillator methods have supported multi-year tests of Lorentz invariance, local position invariance, and fundamental-constant stability \cite{Nagel:2015dd,Tobar2013,Tobar2010}. Arrays of oscillators and cross-correlation methods \cite{XSWisp} may further reject uncorrelated readout noise. The achievable sensitivity is therefore controlled jointly by $Q_p$, pump build-up, coupling, and the measured residual phase-noise spectrum, rather than by $Q_p$ alone.

\section{Conclusions}

Twisted triangular resonators support bulk modes with non-zero complex electric--magnetic overlap and can therefore search for ULDM axions without an external magnetic field. FEM gives $|\mathscr H|\sim1$ for selected linear and ring modes, and the measured resonances of three 3D-printed linear cavities agree well with their predicted twist-angle dependence. The corrected quadrature analysis shows that the real overlap $\mathscr N$ produces AM, whereas the imaginary helicity $\mathscr H$ calculated for these modes produces a reactive frequency perturbation and hence PM. Both measured modulation channels contain the same cavity response factor, but they must be compared with different technical noise spectra. The sensitivity curves remain numerically valid when their assumed noise floor is treated as PM noise. Achieving them therefore requires a dedicated low-phase-noise oscillator, a phase-sensitive readout, and experimental verification of $S_{pm}(f)$ across the axion search band.\\

\noindent\textbf{Acknowledgments}

This work was funded by the Australian Research Council Centre of Excellence for Engineered Quantum Systems, CE170100009 and the Centre of Excellence for Dark Matter Particle Physics, CE200100008.

\providecommand{\noopsort}[1]{}\providecommand{\singleletter}[1]{#1}%


\begin{thebibliography}{10}

\bibitem{Ballon2009}
Douglas~J. Ballon and Henning~U. Voss.
\newblock Classical m\"obius-ring resonators exhibit fermion-boson rotational
  symmetry.
\newblock {\em Phys. Rev. Lett.}, 101:247701, Dec 2008.

\bibitem{Schw51}
Julian Schwinger.
\newblock On gauge invariance and vacuum polarization.
\newblock {\em Phys. Rev.}, 82:664--679, Jun 1951.

\bibitem{Adler69}
Stephen~L. Adler.
\newblock Axial-vector vertex in spinor electrodynamics.
\newblock {\em Phys. Rev.}, 177:2426--2438, Jan 1969.

\bibitem{Gooth2017vg}
Johannes Gooth, Anna~C. Niemann, Tobias Meng, Adolfo~G. Grushin, Karl
  Landsteiner, Bernd Gotsmann, Fabian Menges, Marcus Schmidt, Chandra Shekhar,
  Vicky SuB, Ruben Huhne, Bernd Rellinghaus, Claudia Felser, Binghai Yan, and
  Kornelius Nielsch.
\newblock Experimental signatures of the mixed axial-gravitational anomaly in
  the weyl semimetal nbp.
\newblock {\em Nature}, 547(7663):324--327, 2017.

\bibitem{Ren:2022vd}
Zejian Ren, Dong Liu, Entong Zhao, Chengdong He, Ka~Kwan Pak, Jensen Li, and
  Gyu-Boong Jo.
\newblock Chiral control of quantum states in non-hermitian spin--orbit-coupled
  fermions.
\newblock {\em Nature Physics}, 18(4):385--389, 2022.

\bibitem{Vu:2021wl}
Dung Vu, Wenjuan Zhang, Cuneyt Sahin, Michael~E. Flatte, Nandini Trivedi, and
  Joseph~P. Heremans.
\newblock Thermal chiral anomaly in the magnetic-field-induced ideal weyl phase
  of bi1-xsbx.
\newblock {\em Nature Materials}, 20(11):1525--1531, 2021.

\bibitem{Gooth:2019np}
J.~Gooth, B.~Bradlyn, S.~Honnali, C.~Schindler, N.~Kumar, J.~Noky, Y.~Qi,
  C.~Shekhar, Y.~Sun, Z.~Wang, B.~A. Bernevig, and C.~Felser.
\newblock Axionic charge-density wave in the weyl semimetal (tase4)2i.
\newblock {\em Nature}, 575(7782):315--319, 2019.

\bibitem{Pikulin16}
D.~I. Pikulin, Anffany Chen, and M.~Franz.
\newblock Chiral anomaly from strain-induced gauge fields in dirac and weyl
  semimetals.
\newblock {\em Phys. Rev. X}, 6:041021, Oct 2016.

\bibitem{Wang13}
Zhong Wang and Shou-Cheng Zhang.
\newblock Chiral anomaly, charge density waves, and axion strings from weyl
  semimetals.
\newblock {\em Phys. Rev. B}, 87:161107, Apr 2013.

\bibitem{Wieder:2022um}
Benjamin~J. Wieder, Barry Bradlyn, Jennifer Cano, Zhijun Wang, Maia~G.
  Vergniory, Luis Elcoro, Alexey~A. Soluyanov, Claudia Felser, Titus Neupert,
  Nicolas Regnault, and B.~Andrei Bernevig.
\newblock Topological materials discovery from crystal symmetry.
\newblock {\em Nature Reviews Materials}, 7(3):196--216, 2022.

\bibitem{Fomin:2022aa}
Vladimir~M. Fomin, Roman~O. Rezaev, and Oleksandr~V. Dobrovolskiy.
\newblock Topological transitions in ac/dc-driven superconductor nanotubes.
\newblock {\em Scientific Reports}, 12(1):10069, 2022.

\bibitem{MacKenzie:2021vs}
Lewis~E. MacKenzie and Patrycja Stachelek.
\newblock The twists and turns of chiral chemistry.
\newblock {\em Nature Chemistry}, 13(6):521--522, 2021.

\bibitem{Tang2011}
Yiqiao Tang and Adam~E. Cohen.
\newblock Enhanced enantioselectivity in excitation of chiral molecules by
  superchiral light.
\newblock {\em Science}, 332(6027):333--336, 2011.

\bibitem{Torsi:2008wr}
Luisa Torsi, Gianluca~M. Farinola, Francesco Marinelli, M.~Cristina Tanese,
  Omar~Hassan Omar, Ludovico Valli, Francesco Babudri, Francesco Palmisano,
  P.~Giorgio Zambonin, and Francesco Naso.
\newblock A sensitivity-enhanced field-effect chiral sensor.
\newblock {\em Nature Materials}, 7(5):412--417, 2008.

\bibitem{Hendry:2010ug}
E.~Hendry, T.~Carpy, J.~Johnston, M.~Popland, R.~V. Mikhaylovskiy, A.~J.
  Lapthorn, S.~M. Kelly, L.~D. Barron, N.~Gadegaard, and M.~Kadodwala.
\newblock Ultrasensitive detection and characterization of biomolecules using
  superchiral fields.
\newblock {\em Nature Nanotechnology}, 5(11):783--787, 2010.

\bibitem{Cohen2010}
Yiqiao Tang and Adam~E. Cohen.
\newblock Optical chirality and its interaction with matter.
\newblock {\em Phys. Rev. Lett.}, 104:163901, Apr 2010.

\bibitem{Mun:2020ue}
Jungho Mun, Minkyung Kim, Younghwan Yang, Trevon Badloe, Jincheng Ni, Yang
  Chen, Cheng-Wei Qiu, and Junsuk Rho.
\newblock Electromagnetic chirality: from fundamentals to nontraditional
  chiroptical phenomena.
\newblock {\em Light: Science \& Applications}, 9(1):139, 2020.

\bibitem{Liu:2014ur}
Mingkai Liu, David~A. Powell, Ilya~V. Shadrivov, Mikhail Lapine, and Yuri~S.
  Kivshar.
\newblock Spontaneous chiral symmetry breaking in metamaterials.
\newblock {\em Nature Communications}, 5(1):4441, 2014.

\bibitem{Khanikaev:2013vy}
Alexander~B. Khanikaev, S.~Hossein~Mousavi, Wang-Kong Tse, Mehdi Kargarian,
  Allan~H. MacDonald, and Gennady Shvets.
\newblock Photonic topological insulators.
\newblock {\em Nature Materials}, 12(3):233--239, 2013.

\bibitem{Goryachev2016}
Maxim Goryachev and Michael~E. Tobar.
\newblock Reconfigurable microwave photonic topological insulator.
\newblock {\em Phys. Rev. Applied}, 6:064006, Dec 2016.

\bibitem{Yang16}
Biao Yang, Mark Lawrence, Wenlong Gao, Qinghua Guo, and Shuang Zhang.
\newblock One-way helical electromagnetic wave propagation supported by
  magnetized plasma.
\newblock {\em Scientific Reports}, 6(1):21461, 2016.

\bibitem{Zhang:2020vy}
Xuanru Zhang and Tie~Jun Cui.
\newblock Single-particle dichroism using orbital angular momentum in a
  microwave plasmonic resonator.
\newblock {\em ACS Photonics}, 7(12):3291--3297, 12 2020.

\bibitem{ULACMB2017}
Ren\'ee Hlo\ifmmode~\check{z}\else \v{z}\fi{}ek, David J.~E. Marsh, Daniel
  Grin, Rupert Allison, Jo~Dunkley, and Erminia Calabrese.
\newblock Future cmb tests of dark matter: Ultralight axions and massive
  neutrinos.
\newblock {\em Phys. Rev. D}, 95:123511, Jun 2017.

\bibitem{Marsh17}
Alberto Diez-Tejedor and David J.~E. Marsh.
\newblock Cosmological production of ultralight dark matter axions.
\newblock {\em arXiv:1702.02116 [hep-ph]}, 2017.

\bibitem{Zhang_2018}
Jiajun Zhang, Yue-Lin~Sming Tsai, Jui-Lin Kuo, Kingman Cheung, and Ming-Chung
  Chu.
\newblock Ultralight axion dark matter and its impact on dark halo structure in
  n-body simulations.
\newblock {\em The Astrophysical Journal}, 853(1):51, jan 2018.

\bibitem{Fedderke19}
Michael~A. Fedderke, Peter~W. Graham, and Surjeet Rajendran.
\newblock Axion dark matter detection with cmb polarization.
\newblock {\em Phys. Rev. D}, 100:015040, Jul 2019.

\bibitem{Creedon16}
Daniel~L. Creedon, Maxim Goryachev, Nikita Kostylev, Timothy~B. Sercombe, and
  Michael~E. Tobar.
\newblock A 3d printed superconducting aluminium microwave cavity.
\newblock {\em Applied Physics Letters}, 109(3):032601, 2016.

\bibitem{Romanenko20}
A.~Romanenko, R.~Pilipenko, S.~Zorzetti, D.~Frolov, M.~Awida, S.~Belomestnykh,
  S.~Posen, and A.~Grassellino.
\newblock Three-dimensional superconducting resonators at $t=20$ mk with photon
  lifetimes up to $\ensuremath{\tau}=2$ s.
\newblock {\em Phys. Rev. Applied}, 13:034032, Mar 2020.

\bibitem{Posen20}
S.~Posen, A.~Romanenko, A.~Grassellino, O.S. Melnychuk, and D.A. Sergatskov.
\newblock Ultralow surface resistance via vacuum heat treatment of
  superconducting radio-frequency cavities.
\newblock {\em Phys. Rev. Applied}, 13:014024, Jan 2020.

\bibitem{McA21}
Ben~T. McAllister, Jeremy Bourhill, Wing Him~Jacob Ma, Tim Sercombe, Maxim
  Goryachev, and Michael~E. Tobar.
\newblock Characterization of cryogenic material properties of 3-d-printed
  superconducting niobium using a 3-d lumped element microwave cavity.
\newblock {\em IEEE Transactions on Instrumentation and Measurement}, 70:1--7,
  2021.

\bibitem{PQ1977}
R.~D. Peccei and Helen~R. Quinn.
\newblock Cp conservation in the presence of pseudoparticles.
\newblock {\em Phys. Rev. Lett.}, 38:1440--1443, Jun 1977.

\bibitem{PQ1977b}
R.~D. Peccei and Helen~R. Quinn.
\newblock Constraints imposed by $\mathrm{CP}$ conservation in the presence of
  pseudoparticles.
\newblock {\em Phys. Rev. D}, 16:1791--1797, Sep 1977.

\bibitem{Weinberg1978}
Steven Weinberg.
\newblock A new light boson?
\newblock {\em Phys. Rev. Lett.}, 40:223--226, Jan 1978.

\bibitem{Wilczek1978}
F.~Wilczek.
\newblock Problem of strong $p$ and $t$ invariance in the presence of
  instantons.
\newblock {\em Phys. Rev. Lett.}, 40:279--282, Jan 1978.

\bibitem{Di-Luzio:2021wu}
Luca Di~Luzio, Belen Gavela, Pablo Quilez, and Andreas Ringwald.
\newblock An even lighter qcd axion.
\newblock {\em Journal of High Energy Physics}, 2021(5):184, 2021.

\bibitem{Di_Luzio_2021}
Luca~Di Luzio, Belen Gavela, Pablo Quilez, and Andreas Ringwald.
\newblock Dark matter from an even lighter {QCD} axion: trapped misalignment.
\newblock {\em Journal of Cosmology and Astroparticle Physics}, 2021(10):001,
  oct 2021.

\bibitem{Sokolov:2021uv}
Anton~V. Sokolov and Andreas Ringwald.
\newblock Photophilic hadronic axion from heavy magnetic monopoles.
\newblock {\em Journal of High Energy Physics}, 2021(6):123, 2021.

\bibitem{Visinelli19}
Luca Visinelli and Sunny Vagnozzi.
\newblock Cosmological window onto the string axiverse and the supersymmetry
  breaking scale.
\newblock {\em Phys. Rev. D}, 99:063517, Mar 2019.

\bibitem{Wu19}
Teng Wu, John~W. Blanchard, Gary~P. Centers, Nataniel~L. Figueroa, Antoine
  Garcon, Peter~W. Graham, Derek F.~Jackson Kimball, Surjeet Rajendran,
  Yevgeny~V. Stadnik, Alexander~O. Sushkov, Arne Wickenbrock, and Dmitry
  Budker.
\newblock Search for axionlike dark matter with a liquid-state nuclear spin
  comagnetometer.
\newblock {\em Phys. Rev. Lett.}, 122:191302, May 2019.

\bibitem{Roussy21}
Tanya~S. Roussy, Daniel~A. Palken, William~B. Cairncross, Benjamin~M. Brubaker,
  Daniel~N. Gresh, Matt Grau, Kevin~C. Cossel, Kia~Boon Ng, Yuval Shagam, Yan
  Zhou, Victor~V. Flambaum, Konrad~W. Lehnert, Jun Ye, and Eric~A. Cornell.
\newblock Experimental constraint on axionlike particles over seven orders of
  magnitude in mass.
\newblock {\em Phys. Rev. Lett.}, 126:171301, Apr 2021.

\bibitem{Jiang:2021wn}
Min Jiang, Haowen Su, Antoine Garcon, Xinhua Peng, and Dmitry Budker.
\newblock Search for axion-like dark matter with spin-based amplifiers.
\newblock {\em Nature Physics}, 17(12):1402--1407, 2021.

\bibitem{Bloch22}
Itay~M. Bloch, Gil Ronen, Roy Shaham, Ori Katz, Tomer Volansky, and Or~Katz.
\newblock New constraints on axion-like dark matter using a floquet quantum
  detector.
\newblock {\em Science Advances}, 8(5):eabl8919, 2022.

\bibitem{Goryachev2019}
Maxim Goryachev, Ben~T. McAllister, and Michael~E. Tobar.
\newblock Axion detection with precision frequency metrology.
\newblock {\em Physics of the Dark Universe}, 26:100345, Dec 2019.

\bibitem{Thomson:2021wk}
Catriona Thomson, Maxim Goryachev, Ben~T. McAllister, and Michael~E. Tobar.
\newblock Corrigendum to ``axion detection with precision frequency
  metrology''{$[$}phys. dark universe 26 (2019) 100345{$]$}.
\newblock {\em Physics of the Dark Universe}, 32:100787, 2021.

\bibitem{Lasenby2020}
Robert Lasenby.
\newblock Microwave cavity searches for low-frequency axion dark matter.
\newblock {\em Phys. Rev. D}, 102:015008, Jul 2020.

\bibitem{Lasenby2020b}
Robert Lasenby.
\newblock Parametrics of electromagnetic searches for axion dark matter.
\newblock {\em Phys. Rev. D}, 103:075007, Apr 2021.

\bibitem{berlin2020axion}
Asher Berlin, Raffaele~Tito D'Agnolo, Sebastian~AR Ellis, Christopher Nantista,
  Jeffrey Neilson, Philip Schuster, Sami Tantawi, Natalia Toro, and Kevin Zhou.
\newblock Axion dark matter detection by superconducting resonant frequency
  conversion.
\newblock {\em Journal of High Energy Physics}, 2020(7):1--42, 2020.

\bibitem{Cat21}
Catriona~A. Thomson, Ben~T. McAllister, Maxim Goryachev, Eugene~N. Ivanov, and
  Michael~E. Tobar.
\newblock Upconversion loop oscillator axion detection experiment: A precision
  frequency interferometric axion dark matter search with a cylindrical
  microwave cavity.
\newblock {\em Phys. Rev. Lett.}, 126:081803, Feb 2021.
\newblock [Erratum: Phys. Rev. Lett.127, 019901(2021)].

\bibitem{ABerlin2021}
Asher Berlin, Raffaele~Tito D'Agnolo, Sebastian A.~R. Ellis, and Kevin Zhou.
\newblock Heterodyne broadband detection of axion dark matter.
\newblock {\em Phys. Rev. D}, 104:L111701, Dec 2021.

\bibitem{UpconvCat23}
Catriona~A. Thomson, Maxim Goryachev, Ben~T. McAllister, Eugene~N. Ivanov, Paul
  Altin, and Michael~E. Tobar.
\newblock {Searching for low-mass axions using resonant upconversion}.
\newblock {\em Phys. Rev. D}, 107:112003, Jun 2023.

\bibitem{Cameron_2012}
Robert~P Cameron and Stephen~M Barnett.
\newblock Electric-magnetic symmetry and noether's theorem.
\newblock {\em New Journal of Physics}, 14(12):123019, dec 2012.

\bibitem{Cameron_2012b}
Robert~P Cameron, Stephen~M Barnett, and Alison~M Yao.
\newblock Optical helicity, optical spin and related quantities in
  electromagnetic theory.
\newblock {\em New Journal of Physics}, 14(5):053050, may 2012.

\bibitem{Bliokh_2013}
Konstantin~Y Bliokh, Aleksandr~Y Bekshaev, and Franco Nori.
\newblock Dual electromagnetism: helicity, spin, momentum and angular momentum.
\newblock {\em New Journal of Physics}, 15(3):033026, mar 2013.

\bibitem{Bliokh14}
Konstantin~Y. Bliokh, Yuri~S. Kivshar, and Franco Nori.
\newblock Magnetoelectric effects in local light-matter interactions.
\newblock {\em Phys. Rev. Lett.}, 113:033601, Jul 2014.

\bibitem{Bliokh_2016}
Konstantin~Y Bliokh, Aleksandr~Y Bekshaev, and Franco Nori.
\newblock Corrigendum: Dual electromagnetism: helicity, spin, momentum, and
  angular momentum (2013new j. phys.15033026).
\newblock {\em New Journal of Physics}, 18(8):089503, aug 2016.

\bibitem{BLIOKH20151}
Konstantin~Y. Bliokh and Franco Nori.
\newblock Transverse and longitudinal angular momenta of light.
\newblock {\em Physics Reports}, 592:1--38, 2015.
\newblock Transverse and longitudinal angular momenta of light.

\bibitem{Alpeggiani18}
F.~Alpeggiani, K.~Y. Bliokh, F.~Nori, and L.~Kuipers.
\newblock Electromagnetic helicity in complex media.
\newblock {\em Phys. Rev. Lett.}, 120:243605, Jun 2018.

\bibitem{Wei2020}
Lei Wei and Francisco~J. Rodr\'{\i}guez-Fortu\~no.
\newblock Momentum-space geometric structure of helical evanescent waves and
  its implications on near-field directionality.
\newblock {\em Phys. Rev. Applied}, 13:014008, Jan 2020.

\bibitem{Lipkin:1964wb}
Daniel~M. Lipkin.
\newblock Existence of a new conservation law in electromagnetic theory.
\newblock {\em Journal of Mathematical Physics}, 5(5):696--700, 2022/01/20
  1964.

\bibitem{Peccei2006}
R.~D. Peccei.
\newblock {The Strong CP problem and axions}.
\newblock {\em Lect. Notes Phys.}, 741:3--17, 2008.

\bibitem{electromagneticchirality}
Christophe Caloz and Ari Sihvola.
\newblock Electromagnetic chirality, part 1: The microscopic perspective
  [electromagnetic perspectives].
\newblock {\em IEEE Antennas and Propagation Magazine}, 62(1):58--71, 2020.

\bibitem{1393215}
J.~Krupka, M.E. Tobar, J.G. Hartnett, D.~Cros, and J.-M. Le~Floch.
\newblock Extremely high-q factor dielectric resonators for millimeter-wave
  applications.
\newblock {\em IEEE Transactions on Microwave Theory and Techniques},
  53(2):702--712, 2005.

\bibitem{triangle1}
Ana Mor{\'a}n-L{\'o}pez, Jorge~A. Ruiz-Cruz, Juan C{\'o}rcoles, Jos{\'e}~R.
  Montejo-Garai, and Jes{\'u}s~M. Rebollar.
\newblock Analytical expressions of the q-factor for the complete resonant mode
  spectrum of the equilateral triangular waveguide cavity.
\newblock {\em Electronics Letters}, 55(17):944--947, 2019.

\bibitem{triangle2}
Remi L{\'e}andre, Ana Mor{\'a}n-L{\'o}pez, Juan C{\'o}rcoles, Jorge~A.
  Ruiz-Cruz, Jos{\'e}R. Montejo-Garai, and Jes{\'u}s~M. Rebollar.
\newblock Electromagnetic scattering at the waveguide step between equilateral
  triangular waveguides.
\newblock {\em Advances in Mathematical Physics}, 2016:2974675, 2016.

\bibitem{Sikivie2021}
Pierre Sikivie.
\newblock Invisible axion search methods.
\newblock {\em Rev. Mod. Phys.}, 93:015004, Feb 2021.

\bibitem{Sikivie:1984vr}
P.~Sikivie.
\newblock On the interaction of magnetic monopoles with axionic domain walls.
\newblock {\em Physics Letters B}, 137(5):353--356, 1984.

\bibitem{Wilczek:1987aa}
Frank Wilczek.
\newblock Two applications of axion electrodynamics.
\newblock {\em Physical Review Letters}, 58(18):1799--1802, 05 1987.

\bibitem{tobar2021abraham}
Michael~E. Tobar, Ben~T. McAllister, and Maxim Goryachev.
\newblock Poynting vector controversy in axion modified electrodynamics.
\newblock {\em Phys. Rev. D}, 105:045009, Feb 2022.
\newblock [Erratum: Phys. Rev. D 106, 109903(E) (2022)].

\bibitem{Foster1924}
R.~M. Foster.
\newblock A reactance theorem.
\newblock {\em Bell System Technical Journal}, 3:259--267, 1924.

\bibitem{SokolovMonopole22}
Anton~V. Sokolov and Andreas Ringwald.
\newblock Electromagnetic couplings of axions.
\newblock {\em arXiv:2205.02605 [hep-ph]}, 2022.

\bibitem{sokolov2023generic}
Anton~V. Sokolov and Andreas Ringwald.
\newblock {Generic axion Maxwell equations: path integral approach}.
\newblock {\em arXiv:2303.10170 [hep-ph]}, 2023.

\bibitem{TobarQEMD22}
Michael~E. Tobar, Catriona~A. Thomson, Benjamin~T. McAllister, Maxim Goryachev,
  Anton~V. Sokolov, and Andreas Ringwald.
\newblock {Sensitivity of Resonant Axion Haloscopes to Quantum
  Electromagnetodynamics}.
\newblock {\em Annalen der Physik}, 2200594, 2023.

\bibitem{tobar2023searching}
Michael~E. Tobar, Anton~V. Sokolov, Andreas Ringwald, and Maxim Goryachev.
\newblock Searching for gut-scale qcd axions and monopoles with a high voltage
  capacitor.
\newblock {\em arXiv 2306.13320 hep-ph}, 2023.

\bibitem{Ivanov2009aa}
Eugene~N. Ivanov and Michael~E. Tobar.
\newblock Microwave phase detection at the level $10^-{11}$.
\newblock {\em Review of Scientific Instruments}, 80(4):044701, 2009.

\bibitem{Ivanov2006}
E.N. Ivanov and M.E. Tobar.
\newblock Low phase-noise microwave oscillators with interferometric signal
  processing.
\newblock {\em {IEEE} Transactions on Microwave Theory and Techniques},
  54(8):3284--3294, aug 2006.

\bibitem{Ivanov2009}
E.~N. Ivanov and M.~E. Tobar.
\newblock Low phase-noise sapphire crystal microwave oscillators: current
  status.
\newblock 56:263--269, 2009.

\bibitem{Ivanov1998}
E.~N. {Ivanov}, M.~E. {Tobar}, and R.~A. {Woode}.
\newblock Microwave interferometry: application to precision measurements and
  noise reduction techniques.
\newblock {\em and Frequency Control IEEE Transactions on Ultrasonics,
  Ferroelectrics}, 45(6):1526--1536, November 1998.

\bibitem{locke}
C.~R. Locke, E.~N. Ivanov, J.~G. Hartnett, P.~L. Stanwix, and M.~E. Tobar.
\newblock Invited article: Design techniques and noise properties of
  ultrastable cryogenically cooled sapphire-dielectric resonator oscillators.
\newblock {\em Review of Scientific Instruments}, 79(5):051301, 2008.

\bibitem{rubiolaAM}
E.~Rubiola and R.~Brendel.
\newblock The am noise mechanism in oscillators.
\newblock {\em IEEE IFCS Conf. Proc.}, pages 33--39, 2009.

\bibitem{Nagel:2015dd}
Moritz Nagel, Stephen~R. Parker, Evgeny~V. Kovalchuk, Paul~L. Stanwix, John~G.
  Hartnett, Eugene~N. Ivanov, Achim Peters, and Michael~E. Tobar.
\newblock Direct terrestrial test of lorentz symmetry in electrodynamics to
  10$^{-18}$.
\newblock {\em Nature Communications}, 6:8174 EP, 09 2015.

\bibitem{Tobar2013}
M.~E. Tobar, P.~L. Stanwix, J.~J. McFerran, J.~Guena, M.~Abgrall, S.~Bize,
  A.~Clairon, Ph. Laurent, P.~Rosenbusch, D.~Rovera, and G.~Santarelli.
\newblock Testing local position and fundamental constant invariance due to
  periodic gravitational and boost using long-term comparison of the {SYRTE}
  atomic fountains and h-masers.
\newblock {\em Physical Review D}, 87(12), jun 2013.

\bibitem{Tobar2010}
Michael~Edmund Tobar, Peter Wolf, Sebastien Bize, Giorgio Santarelli, and
  Victor Flambaum.
\newblock Testing local lorentz and position invariance and variation of
  fundamental constants by searching the derivative of the comparison frequency
  between a cryogenic sapphire oscillator and hydrogen maser.
\newblock {\em Physical Review D}, 81(2), jan 2010.

\bibitem{XSWisp}
Ben~T. Mcallister, Stephen~R. Parker, Eugene~N. Ivanov, and Michael~E. Tobar.
\newblock Cross-correlation signal processing for axion and wisp dark matter
  searches.
\newblock {\em IEEE Transactions on Ultrasonics, Ferroelectrics, and Frequency
  Control}, 66(1):236--243, 2019.

\end{thebibliography}
\end{document}